\documentclass[runningheads]{llncs}

\usepackage{geometry}
\usepackage{booktabs} 
\usepackage{todonotes}
\usepackage{listings}
\usepackage{multicol}
\usepackage{multirow}
\usepackage{enumerate}
\usepackage{balance}
\usepackage[font=scriptsize]{subfig}
\usepackage{graphicx}

\usepackage[utf8]{inputenc}
\usepackage{amsmath}
\usepackage{amssymb}
\usepackage{amsfonts}
\usepackage{hyperref}

\begin{document}

\definecolor{myblue}{HTML}{3366FF}
\definecolor{editorGray}{rgb}{0.95, 0.95, 0.95}
\definecolor{editorOcher}{rgb}{1, 0.5, 0} 
\definecolor{editorGreen}{rgb}{0, 0.5, 0} 
\lstdefinelanguage{HTML5}{
        language=html,
        sensitive=true, 
        alsoletter={<>=-},
        otherkeywords={
        <html>, <head>, <title>, </title>, <meta, />, </head>, <body, <figure, <article, <ol, <input, <u,
        <canvas, \/canvas>, <script>, </script>, </body>, </html>, <!, <style>, </style>, ><, <h2, <form,
        <ul, <\/h2>, <\/form>, <\/ul>, <\/figure>, <\/article>, <\/ol>, <\/input>, <\/u>, </meta>
        },  
        ndkeywords={
        id=, style=, spellcheck=, dir=, title=, lang=, translate=, contenteditable=, tabindex=, class=, size=, type=,
        name=, formnovalidate, http-equiv=, content=
        },  
        morecomment=[s]{<!--}{-->},
        tag=[s]
}

\lstset{%
    basicstyle={\footnotesize\ttfamily},   
    frame=l,
    xleftmargin={0.75cm},
    numbers=left,
    stepnumber=1,
    firstnumber=1,
    numberfirstline=true,
    keywordstyle=\color{myblue}\bfseries,
    commentstyle=\color{darkgray}\ttfamily,
    ndkeywordstyle=\color{editorGreen}\bfseries,
    stringstyle=\color{editorOcher},
    language=HTML5,
    alsodigit={.:;},
    tabsize=2,
    showtabs=false,
    showspaces=false,
    showstringspaces=false,
    extendedchars=true,
    breaklines=true,        
    literate=%
    {Ö}{{\"O}}1
    {Ä}{{\"A}}1
    {Ü}{{\"U}}1
    {ß}{{\ss}}1
    {ü}{{\"u}}1
    {ä}{{\"a}}1
    {ö}{{\"o}}1
}

\setlength{\belowcaptionskip}{-10pt}

\title{Recurrent Neural Networks for Fuzz Testing Web Browsers}

 \author{Martin Sablotny\orcidID{0000-0002-9836-8254} \and
 Bj\o rn Sand Jensen \and \\
 Chris W. Johnson}
 \authorrunning{M. Sablotny et al.}
 \institute{University of Glasgow, School of Computing Science, Glasgow, Scotland\\
 \email{m.sablotny.1@research.gla.ac.uk, bjorn.jensen@glasgow.ac.uk, christopher.johnson@glasgow.ac.uk}}

\maketitle
\begin{abstract}
Generation-based fuzzing is a software testing approach which is able to discover different types of bugs and vulnerabilities in software. 
It is, however, known to be very time consuming to design and fine tune classical fuzzers to achieve acceptable coverage, even for small-scale 
software systems. To address this issue, we investigate a machine learning-based approach to fuzz testing in which we outline a family of test-case generators 
based on Recurrent Neural Networks (RNNs) and train those on readily available datasets with a minimum of human fine tuning. The proposed 
generators do, in contrast to previous work, not rely on heuristic sampling strategies but principled sampling from the predictive distributions.
We provide a detailed analysis to demonstrate the characteristics and efficacy of the proposed generators in a challenging web browser testing scenario. The empirical results show that the RNN-based generators are able to provide better coverage than a mutation based method and are able to discover 
paths not discovered by a classical fuzzer. Our results supplement findings in other domains suggesting that generation based fuzzing with RNNs is a viable route to better software quality conditioned on the use of a suitable model selection/analysis procedure.

\end{abstract}

\section{Introduction}
Fuzz testing has recently enjoyed increased popularity in theoretical and practical software testing. 
This can be primarily attributed to the apparent capability to trigger unintended behaviour in complex
software systems, e.g. the summary of bugs found by American Fuzzy Lop (AFL) \cite{afl} and further 
evidenced by the use of fuzz testing in software companies like Microsoft and Google (e.g. through 
their open-source tool ClusterFuzz \cite{google_clusterfuzz}) which shows success and applicability 
in many different domains. However, the standard approach of combining mutation on a set of input 
examples with an evolutionary approach has its limitation with increasing necessity of keywords and 
compliance to syntactic rules (e.g. HTML as considered in this work). Those problems can be tackled 
by generation-based fuzzers that are able to comply to those rules, use the correct keywords and 
generate novel inputs. Traditionally, the time needed to develop generation-based fuzzers is dependent 
on the input specification's complexity. For example it is less time consuming to develop a generator 
for a network protocol, which has a single field with three different possible values compared to 
implementing the File Transfer Protocol (FTP) \cite{ftp_rfc} with it various fields and states. 
In addition, it is necessary to find the right balance between introduced errors and overall correctness 
to trigger code paths that lead to unintended behaviour.

The main bottleneck in the development of generation-based fuzzers is the need for a strict understanding and 
implementation of the input file format. Therefore, the potentially complex input specification has to be studied 
carefully to transfer it into a test case generator, which then needs to be fine tuned in order to find the right 
balance between correctness and introduced errors into the test cases. This implicit optimization process looks 
to maximize code coverage by generating test cases that deviate in certain areas from the given specification 
and therefore are capable of exercising different low-level execution paths. Thus, it is clear that methods 
which could automatically derive or lean the input specification would be able to speed up software testing 
by faster deployment of generation-based fuzzing techniques. This would potentially lead to an increase in 
software security and stability. 

Learning an input specification (e.g. syntactic rules) is obviously not trivial, especially due to the long time
dependencies input specifications can apply. Those dependencies have an direct impact on the possible outputs
at a certain position and therefore have to be captured by a learning algorithm to produce specification adhering
outputs. However, recent advancements in generative machine learning models (\cite{sutskever2011generating},
\cite{balog2016deepcoder}, \cite{cho2014learning}, \cite{bahdanau2014neural}) have demonstrated how machine 
learning models can be use to learn complex rules and distributions from examples and generate new examples 
from acquired knowledge. 

These advancements have been previously explored for fuzz testing
by Godefroid \textit{et. al.} \cite{godefroid2017learn}. They demonstrated the use of deep neural networks to
generate PDF-objects, which were used as input for a rendering engine. Those input files were able to trigger
new instructions in the rendering engine. However, they focused on the tension between learning the correct 
input structure and fuzzing - or in other words, finding the balance between adhering to the learned 
specification and deviating from it. They did not provide an analysis of the learning process itself and gave 
no comparison to a naive mutation based baseline. In addition, they have not provided any information about the
overlap between the baseline and their proposed sampling strategies. In order to use deep
learning models during fuzz testing, it is important to see whether it is worth the development and training.
Therefore, it is necessary to compare it with an easy to implement approach, like a naive mutation algorithm. 
The analysis of an existing overlap between different approaches also gives more insight into the model and 
sampling choice, since it is important to trigger as much new execution paths as possible during testing to 
find the ones that trigger unintended behaviour. 

In this work, we investigate how Recurrent Neural Networks (RNNs) with different types of cells can be trained 
and used as a HTML-fuzzers. 
The models are trained on a dataset created by a generation based HTML-fuzzer, which allowed us to adjust the 
dataset size and complexity in a fast and systematic way. We use the models to generate new HTML-tags from the 
resulting probability distribution, which were used to form test cases. Those were executed with Firefox 
\cite{moz_ff} to gather their code coverage data and compared to a baseline generated by the HTML-tags from 
the dataset and a naive mutated dataset. Thus, the contribution of the paper includes:
\begin{enumerate}[-] 
\item[-] A systematic and robust approach for training and evaluating recurrent neural networks with different types of cells for HTML fuzz testing.
\item[-] A procedure and metrics for model-selection and comparison of machine learning fuzzers against standard and a vanilla mutation-based methods including a similarity-based analysis.
\item[-] An extensive empirical evaluation on a web browser, demonstrating that learned fuzzers are able to outperform standard test methodologies. 
\item[-] Open-source implementation and data available via Github \footnote{Code and data is available from \url{https://github.com/susperius/icisc_rnnfuzz}}. 
\end{enumerate}

\section{Background}
\subsection{Fuzzing}
\begin{figure*}[t!]
	\centering
	\includegraphics[width=.9\textwidth]{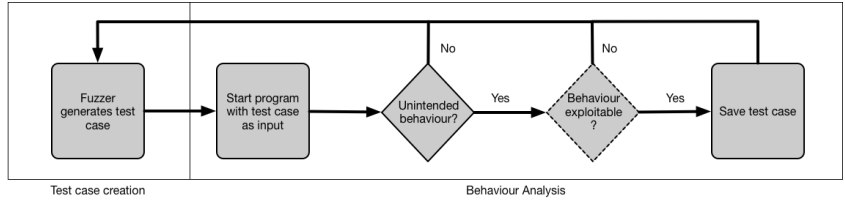}
	\caption{Classic Fuzzing Workflow for finding security related flaws}
	\label{fig:fuz}
\end{figure*}
Fuzz testing is a dynamic software testing approach, hereby dynamic means the software under test is actually executed in contrast to statically analysed. The goal of the fuzz test is to provoke unintended behaviour that was not  detected in earlier testing stages, therefore software under test is executed with inputs created by a so-called fuzzer.
Those inputs do not fully comply with the underlying input specification in order to find paths that lead to a state that 
triggers unintended behaviour. 
We adopt a broad definition of unintended behaviour, which makes it applicable for various kinds of software and devices
\cite{sutton2007fuzzing}.
For example, during fuzz testing desktop software, unintended behaviour can be the termination of a running process or 
even the possibility to take control over a process. Whereas during the test of a web application unintended behaviour 
might be defined as an information leak or the circumvention of access restriction both cases might happen due to a 
SQL-injection vulnerability, where arbitrary input is used as a valid SQL-statement.

As those examples highlight, a case of unintended behaviour becomes more severe if it could provide an attacker with 
an advantage. Here advantage can mean everything from accessing restricted information to taking over control of a 
device.
In order to find those vulnerabilities fuzz testing is utilised. The general workflow during fuzz testing is 
shown in \autoref{fig:fuz}. The testing itself is split in two parts first the test case generation and secondly 
the behaviour analysis.
In general, the creation of test cases during fuzzing can be divided into the two categories: mutation based and 
generation based \cite{sutton2007fuzzing}, \cite{demott2006evolving} and \cite{oehlert2005violating}. First mutation 
based fuzzing uses a valid input set and a mutation fuzzing in order to derive new test cases from the input set. 
This type of fuzzing can be implemented quickly if the input examples are available (e.g. JPG files). The main 
disadvantage is that test cases created by plain mutation based fuzzing are not able to quickly discover code 
paths deep in the call tree because many created test cases are filtered out in early program execution stages. A very prominent 
and successful example of this category is the aforementioned fuzzer AFL with its evolutionary mutation approach.
Secondly, generation based fuzzing uses an approach where test cases are created from scratch, for example through 
grammar based creation. This method needs a lot of effort during studying the input structure and developing the 
generator but in general it is able to discover deeper lying code paths. However, a balance between complying to the 
rules and breaking them has to be found in order to provoke unintended behaviour in the target.

\subsection{Recurrent Neural Networks}
\label{sec:rnn}
The input data for many software products is readily available on the internet (e.g. HTML, JPG, PNG) and
deep learning algorithms have shown their performance in different use cases especially where they are
trained on a large available dataset, for example text generation \cite{sutskever2011generating},
program creation \cite{balog2016deepcoder} and machine translation \cite{cho2014learning}, 
\cite{bahdanau2014neural}. This led us to the use of a generative model for the test case creation during 
fuzz testing. In addition the structure of HTML and other input formats, where the actual character or byte
is dependant on the previous positions in a sequence led to the use of RNNs.

RNNs are used to model sequential data, e.g. for text generation \cite{sutskever2011generating}, language
modelling and music prediction \cite{pascanu2013construct}. They use a hidden state as short term memory
which carries information between time steps. The conventional RNN with input $\mathbf{x_t}$ is defined through a 
hidden state vector $\mathbf{h_t}$ and an output $\mathbf{\hat{y}_t}$  at time step $t$ as follows
\[ \mathbf{h_t} = f_h(\mathbf{x_t}, \mathbf{h_{t-1}}) \,\,\, , \,\,\, \mathbf{\hat{y}_t} = f_o(\mathbf{h_t}), \]
with $f_h$ and $f_o$ being the hidden transformation and output function respectively. Hereby, the input 
$\mathbf{x_t}$ can be a $N$-dimensional vector, representing the input structure, e.g. a single pixel's RGB
values at position $t$.

As described by Hochreiter \cite{hochreiter1991untersuchungen} and later by
Bengio et al. \cite{bengio1994learning}, RNNs suffer from either the vanishing or exploding gradient
problem. This means that the weight updates are becoming infinitesimal during training,
which consumes a lot of time but does not lead to a better optimised network.
Hochreiter and Schmidhuber introduced the concept of Long-Short Term Memory (LSTM) cells \cite{hochreiter1997long}
RNNs using those cells do not suffer from the vanishing (exploding) gradient problem. LSTM cells use a hidden
state, a candidate value and three gates namely a forget gate, an input gate
and an output gate. The gates control how much information is forgotten, used from the input and
controlling the flow into the new hidden state respectively. They are default feed forward neural networks 
and each have their own trainable parameters.

Another popular RNN cell, the Gated Recurrent Unit (GRU) was introduced by Cho et al. \cite{cho2014learning}.
This unit only uses two gates, a reset and an update gate. Here the reset gate controls what information from
the past hidden state is forgotten and the update gate controls the information flow into the new hidden
state. This simpler model arguably makes it easier to train than a standard LSTM based model.

The capability to learn sequential structures, where dependencies to former inputs exist, is obviously an 
important characteristic when learning input format structures for test case generation. This is especially evident in for example HTML where there are long term dependencies between an opening-tag and the corresponding closing-tag.

\section{Stacked RNN for HTML-Fuzzing}
\begin{figure*}[!t]
	\centering
	\includegraphics[width=0.95\textwidth]{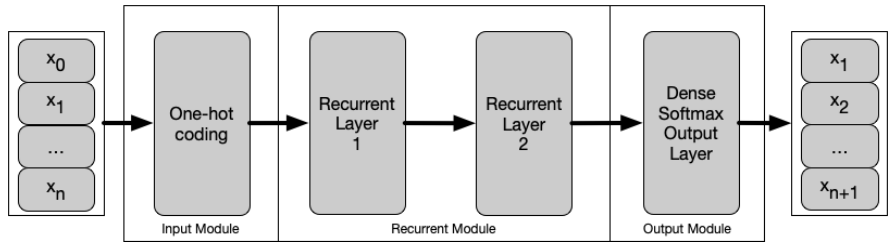}
	\vspace{-0.1cm}
    \caption{Model overview for a stacked RNN with 2 recurrent layers (either LSTM or GRU)}	
	\label{fig_model}
\end{figure*}
The basic concept of the model used in this work is shown in Figure \ref{fig_model}. The model 
consists of three modules. First, the input module,
let $X = \{x_1, x_2, \hdots x_N\}$ be the sequence of input values with 
$ x_t \in \mathbb N_0 \ |\ 1 \leq t \leq N $,
where $x_t$ is the natural number representing the character at position $t$ in the input sequence.
For example the 
character 'f' is at position $t$ in the input sequence, its assigned number is 17 and $x_t=17$.

The input module then takes such a $x_t$ and transforms it into a one-hot coded 
vector
$\mathbf{\hat{x}_{t} \in \mathbb R^I}$
with $I = max(X) + 1$, the one is added to account for the zero. Let $\mathbf{\hat{x}_{t}} = (\hat{x}_1, \hat{x}
_2, \hdots, \hat{x}_I)^\intercal $ then
\[ \hat{x}_j = 0 \ \forall\ 1 \leq j \leq I\ :\ j \not = x_t\ \lor\ \hat{x}_j = 1\ \Leftrightarrow\ j = x_t , \] 
and for the former example character 'f' all $\hat{x}_j=0$, except for $\hat{x}_17$, which equals $1$. This conversion from integer values is necessary as interpret our input as categorical data 
(each character is its own category) and those categories are handled as features during the training process.

Secondly, the recurrent module consists of LSTM or GRU nodes as described in Section \ref{sec:rnn} with
$s,l \in \mathbb N$
hereby $s$ is the internal size of the nodes and $l$ the amount of layers used, e.g. $l=2$ for the LSTM based
model shown Figure \ref{fig_model}.\\
LSTM cells have demonstrated a high performance gain compared to the basic RNN approach as demonstrated by
Chung et al. \cite{chung2014empirical}. Gated Recurrent Units (GRUs) introduced by 
Cho et al.\cite{cho2014learning} perform similar to LSTM cells \cite{chung2014empirical}, however Jozefowicz
et al. \cite{jozefowicz2015empirical} have shown that LSTM cells perform better during XML modelling. We
decided to evaluate the performance of both cells to analysis whether the XML modelling results are
transferable to HTML modelling.

Finally, the output layer consists of a default feed forward network with $I$ nodes.
It takes the output of the last recurrent layer $\mathbf{h}^l_t \in \mathbb R^s$ as input value and
after computing its output the $softmax$ function is applied.
The resulting $\mathbf{\hat{y}_t}$ provides the probability distribution for predicting the next value of the input sequence. The goal during training is to minimise the cross entropy loss function
\[
\mathcal L(\Theta) = -\frac{1}{N}\sum\limits_{i=1}^{N} \mathbf{y_i}\log(\mathbf{\hat{y}_i)}+(1-\mathbf{y_i})\log(1-\mathbf{\hat{y}_i}),
\] 
where $\Theta$ denotes the model's parameters (i.e. a collection of $\mathbf{W}$'s and $\mathbf{b}$'s). In order to find a $\Theta$ that minimises the above loss $\mathcal L$ the ADAM \cite{kingma2014adam} optimisation algorithm is applied. It is a gradient-based optimisation algorithm which only needs first order gradients and has a reduced memory footprint compared to other algorithms. Additionally, Dropout (30\% dropout probability) \cite{srivastava2014dropout} is used as regularisation.

\section{Experiments}
The following sections present the methodology that was used to validate our application of RNNs to generate test cases 
for fuzz testing of cyber security in complex systems. 

The basic idea is to train the aforementioned  neural networks with different depths on a large collection of HTML-tags. 
After training those models are used to generate HTML-tags directly using the probability distribution over characters 
given the sequence. The generated output is then used as input for a web browser. This browser is instrumented in order to
gather the code coverage data during execution on a basic blocks basis. 
The collected code coverage data is then used to compare the models' performances with code coverage data 
collected by executing the dataset's HTML-tags and a naive mutation strategy performed on this HTML-tags. 

\subsection{Environmental Setup and Implementation}
The model training took place on a Ubuntu 16.04 system equipped with a single NVIDIA GeForce 1080 Ti and a NVIDIA GeForce 
TITAN Xp, which shortens the necessary training time by utilising their parallel computational capabilities. The models  
were implemented using Google's TensorFlow framework \cite{tensorflow2015-whitepaper} along with its Python bindings. 
This frameworks already provides the necessary cell types, optimisation algorithm and loss function for our model, which 
shortens the development time. 

The code coverage data was collected on a Virtual Machine (VM) also running Ubuntu 16.04 and Firefox 57.0.1, which allows 
to run in so-called headless mode. In this mode Firefox does not display the graphical user interface, but it still renders 
the webpage. We also modified the standard configuration in order to disable internal services to avoid as much false code 
coverage data as possible. Furthermore safe mode was disabled, because during the automated code coverage collection Firefox
was not closed correctly and therefore might tries to start in safe mode after just a few test cases.
The use of the headless mode also saves time during the code coverage collection, which was collected by DynamoRIO's drcov tool 
(see \autoref{sec_data_coll}). The VM itself utilises 16 GB of RAM and a Solid State Disk. A VM was used to facilitate parallel 
data collection via cloning and deploying onto multiple host systems.
\subsection{Data Set Generation}
\label{sec:exp:ds_gen}
In order to provide a reproducible and controlled experiment, the training (and ground-truth) data set was 
generated by an existing HTML-fuzzer included in PyFuzz2 \cite{sablotny_pyfuzz2}. It provides a controllable
generator thus ensuring less uncertainty about the variation within the training dataset in comparison to
collecting a dataset from the Internet. Therefore, it was possible to control the complexity of the generated
HTML on a per tag basis, whereas a collected set would have to be parsed and then filtered for unwanted 
HTML-tags to control the resulting dataset.

\begin{lstlisting}[caption={Example from the training set.},float=bp,floatplacement=bp, captionpos=b, label={lst_simple_set}]
<h2 id="id0" style="style" spellcheck="false" dir="rtl" title="eval(n1, $)"> 2e100 </h2>
<ul id="id3" style="style" translate="no" contenteditable="true" tabindex="4400000000"> 4400000000 </ul>
\end{lstlisting}

The pre-existing fuzzer was modified in order to avoid nesting of HTML tags, remove all Cascading Style Sheets 
and output exactly one HTML tag per line. Due to the restriction of not having nested HTML-tags some like $td$ 
or $th$ are excluded because they need an outer tag in this example $table$. Those restrictions were introduced 
to reduce to focus on the fundamental problem by reducing the overall data set complexity. This further reduced 
the necessary model complexity and effectively the time needed to train those models.

Listing \ref{lst_simple_set} shows an excerpt from the data set used for training the models, which highlights 
the modification mentioned above. The created file consisted of 409,000 HTML-tags, which results in a total size 
of 36MB.

\subsection{Training}
All models were trained to predict the input shifted by one on a per character basis. For example take 
"$<h2\ i$" from line 1 in Listing \ref{lst_simple_set} as input sequence of length 5 then the label for 
that particular  input sequence would be "$h2\ id$". The actual sequence length used during training 
was 150 characters and each model was trained for 50 epochs, which has shown sufficient for the models 
to converge. In order to train the models we used the previously mentioned ADAM 
\cite{kingma2014adam} optimisation algorithm. The starting learning rate was set to $0.001$ and halved 
every 10 epochs. The models were trained with a batch size of 512. The internal size of the LSTM and
GRU cells  was set to 256 for all models trained and the number of layers varied from 1 to 6. The weights 
of the layer were initialised by the Glorot uniform initializer \cite{glorot2010understanding}. So the 
weights are drawn from a uniform distribution in the interval 
$(- \frac{\sqrt{6}}{\sqrt{n_j + n_{j+1}}},\ \frac{\sqrt{6}}{\sqrt{n_j + n_{j+1}}})$, 
with $n_j$ being the internal size of layer $j$.

The first 30MB of the data set were used for training and an additional generated 1MB for validation. 
All models were trained on 5 different training/validation splits repeated 3 times with different 
initialization (to mitigate extremely poor local minima) which results in a total of 90 trained models
per cell type. The splits were chosen randomly without overlapping parts.

\subsection{Data Collection}
\label{sec_data_coll}
The code coverage data was collected by executing Firefox instrumented by DynamoRIO's $drcov$ \cite{dynamoRIO}. This
tool gathers data about the executed basic blocks of the program under test. The collected code coverage data 
was parsed for uniquely executed basic blocks inside of Firefox's \verb+libxul.so+ library, which includes
the whole web engine responsible for HTML rendering. It is possible to identify those basic blocks even when
the process is restarted because the recorded data uses the offset of the basic block from the base address of 
the library in memory and this offset is always the same for a fixed version. Hereby a basic block is defined
as a linear sequence of machine instructions with a single entry (branch target) and single exit (branch instruction).

All test cases consisted of a basic HTML-template with the HTML-tags inserted into the body tag. Initial experiments showed
that executing the same test case multiple times returns different code coverage data. This is due to the other
functions that are bundled into the \verb+libxul.so+ library, which are not part of the web engine itself. Those functions might for example only be executed after a number of restarts or in fixed time intervals. In order to identify the 
corresponding basic blocks the blank HTML-template was executed $1,024$ times and the resulting code coverage 
was store for later use.

The comparison baseline was established by using the HTML fuzzer to create $6\times 16,384$ HTML-tags 
Each collection of $16,384$ HTML-tags was then used to create two datasets, one containing $64$ files with $256$ 
HTML-tags each  and a second one with $128$ files containing $128$ HTML-tags. This resulted in twelve datasets.

In order to establish a second baseline for comparison, additional test sets were created by mutating the
dataset test cases and collecting the code coverage from those. A simple mutation function was applied with a fixed chance that a position is replaced by a randomly chosen character (only characters 
that were already present in the dataset).
The results were $20$ additional test case sets, $10$ sets consisting of $128$ cases with $128$ HTML-tags each 
and $10$ sets consisting of $64$ cases with $256$ HTML-tags each, resulting in a total of $1,920$ additional cases. 
The replacement probability varied between $0.1\%$ and $51.2\%$. This was done to ensure that there is difference 
and therefore an incentive to use a trained model for test case creation instead of implementing a naive mutation 
based approach.

For each trained model, a total $16,384$ HTML-tags were generated and then used to create two different sets 
of test cases. The first set used $128$ HTML-tags per case, which resulted in $128$ cases per model trained,
whereas the second set used $256$ HTML-tags per case, which resulted in $64$ cases per model.
This was done to analyse the impact of HTML-tags on code coverage and to observe the relationship 
with the model performance. 
\begin{lstlisting}[caption={Example HTML-tag from a 1-layer LSTM model}, float=bp,floatplacement=bp, captionpos=b, label={res:lst:one}]
<war id="id55804" scellcheck="false" tpalleaeck="false" class="style_class_0" title="50000000"> null</sab>
\end{lstlisting}
The HTML-tags were generated by using the "$<$" character as starting input, sampling the next character
from the resulting probability distribution, which was then used as new input. This was repeated until a "\textbackslash{n}"
(newline character) was sampled, since it marks the end of a HTML-tag.

Finally, the set difference between the collections of basic block sets from the test cases and the blank cases was 
computed to filter out the aforementioned irrelevant basic blocks.

\subsection{Results}\label{sec:res}

\begin{figure}[t!]
	\centering
	\subfloat[LSTM\label{fig_valloss_lstm}]{\includegraphics[width=.5\textwidth]{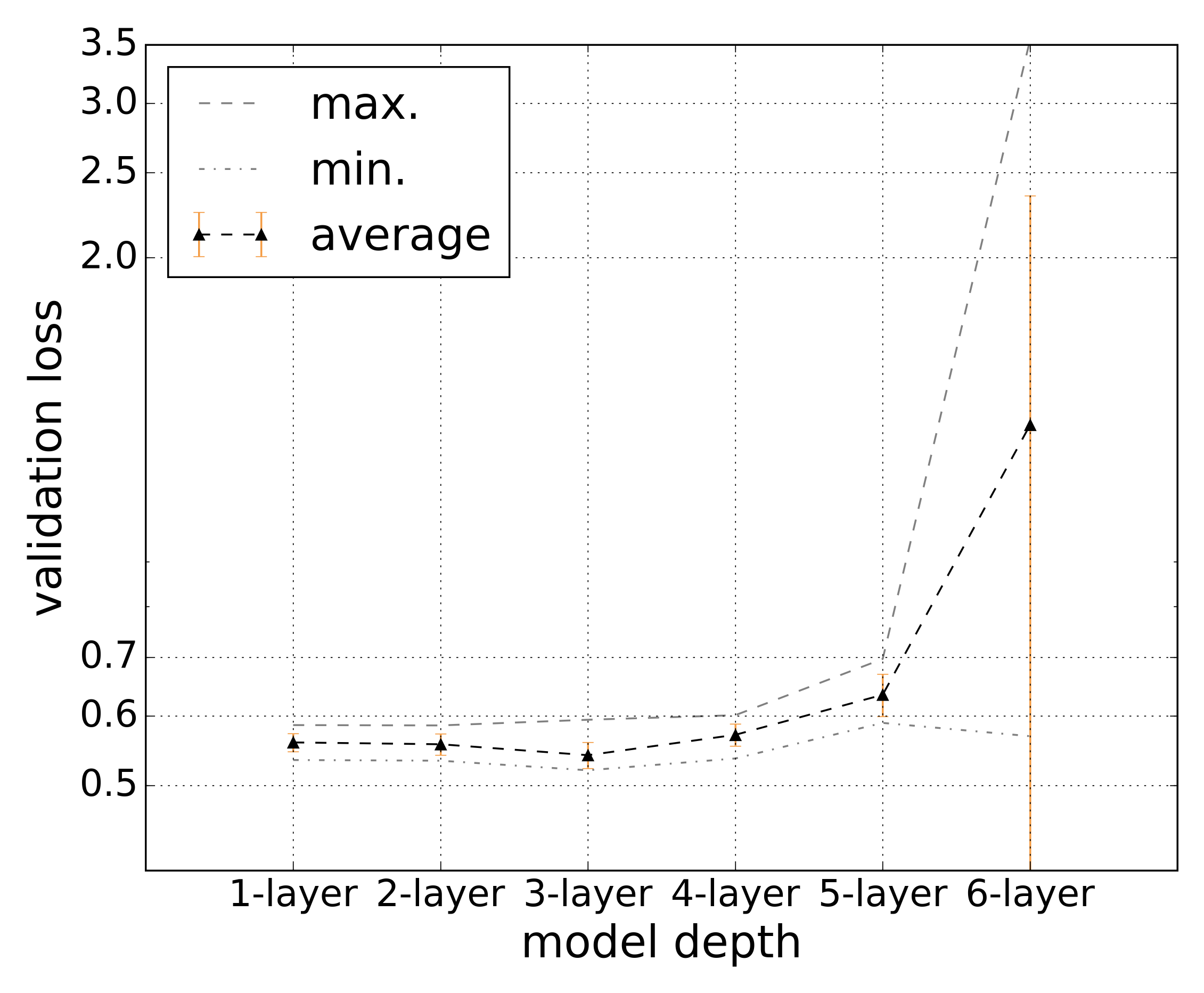}}
	\subfloat[GRU\label{fig_valloss_gru}]{\includegraphics[width=.5\textwidth]{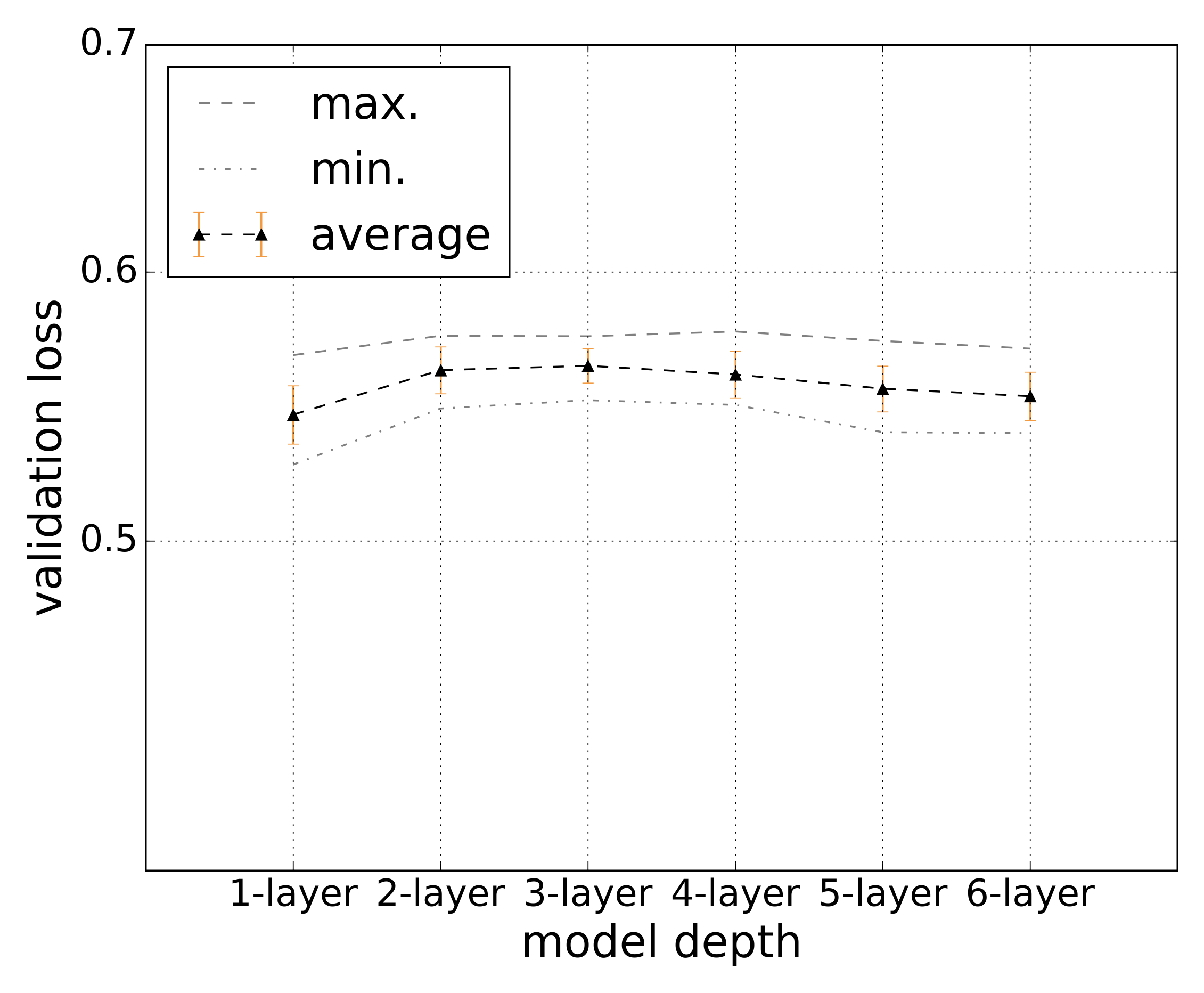}}
	\caption{Average validation loss for models of different complexity (i.e. number of layers) models and dataset splits. Error-bars indicate the standard deviation.}   
	\label{fig_valloss}
\end{figure}

The training phase already showed a difference in behaviour between the two cell types. The LSTM based models
showed a decrease in average validation loss and standard deviation up to three layers, as shown in Figure
\autoref{fig_valloss_lstm}, with an increase afterwards. Especially, the 6-layer models show a large standard
deviation and a huge increase in average validation loss compared to the other models This indicates that those
models have too many parameters in order to be trained on our problem and training set. This behaviour is to be 
expected from a general machine learning perspective and since the training process is the same compared to other 
similar applications using generative neural networks, like generating text.

In contrast the training of the GRU based models showed a small increase from the 1-layer
models to the 2-layers case, but a decrease afterwards with overall small differences in the standard deviation.
This indicates that the GRU based models are either better suited to reproduced the input structure or do not 
reach the overall complexity of the 6-layer LSTM based model, which is also supported by comparing the trainable 
parameters of those models. The GRU based model has $2,276,971$ compared to $3,026,795$.

\begin{lstlisting}[caption={Example HTML-tag form a 3-layer LSTM model},float=bp,floatplacement=bp,  captionpos=b, label={res:lst:three}]
<p id="id38564" lang="mk"> BBBBBBBBBBBBBBBBBBBBBBBBBBBBBBBBBBBBBBBB</p>
<head id="id240801" sang="al" style="style" class="style_class_0" dir="rtl"> 7500000000</pre>
\end{lstlisting}

\begin{figure}[th!]
	\centering
	\includegraphics[width=.5\textwidth]{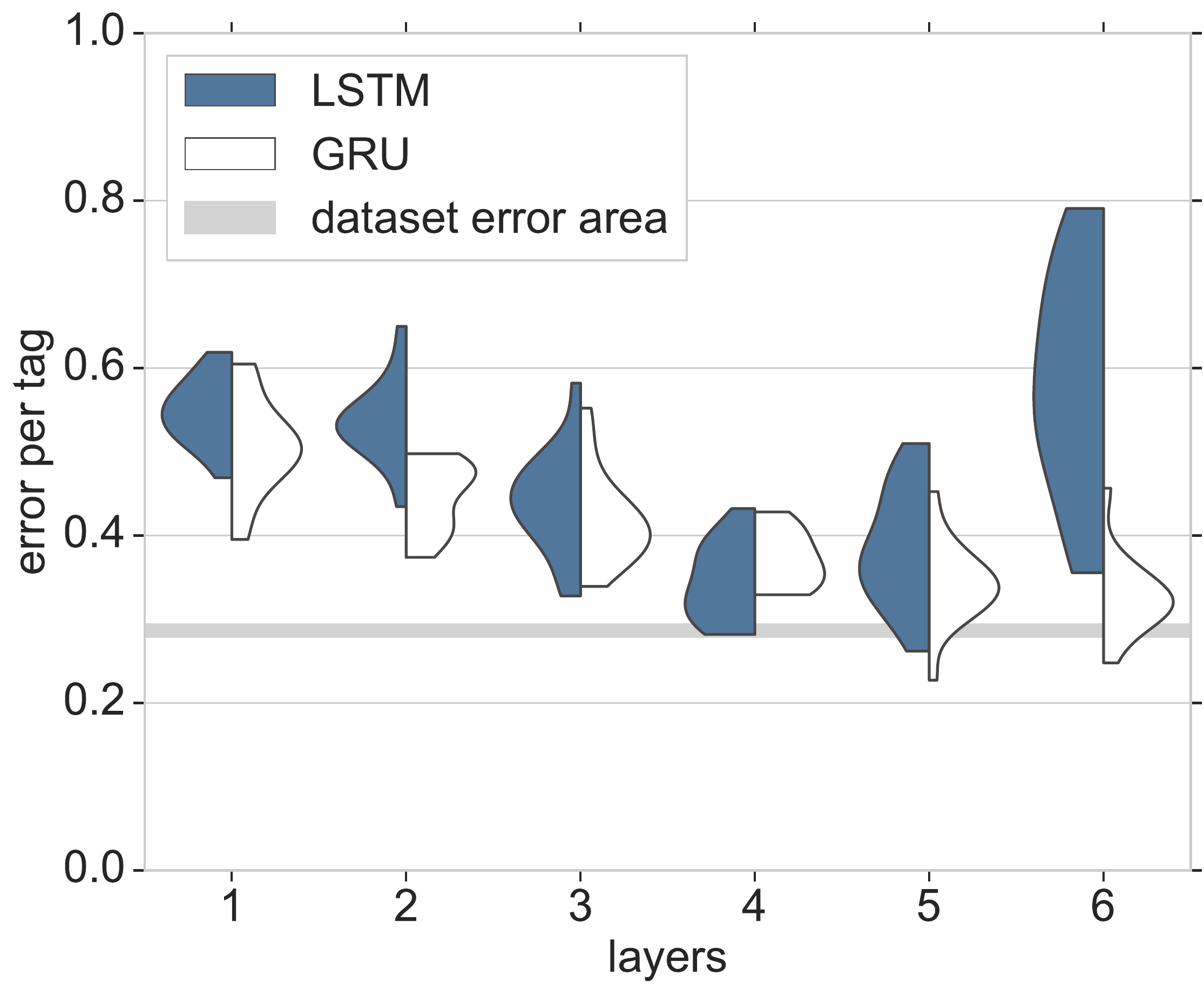}
	\caption{Average error rate per HTML-tag generated by the LSTM and GRU based model in comparison to the datasets.}
	\label{fig_html_err}
\end{figure}
Overall, a small numeric difference in validation loss can lead to a big difference in the quality of the resulting HTML-tags.
For example \autoref{res:lst:one} shows an excerpt generated by a 1-layer LSTM model. It is barely recognisable as 
HTML and the model did not generate existing HTML-opening and closing tags and two of the generated 
HTML-attributes are misspelled in this particular example. In contrast to that \autoref{res:lst:three} 
shows two HTML-tags generated by a 3-layer LSTM model. Both use only existing HTML-tags, however the second one 
does not use the correct closing tag and misspelled one attribute name.
Further evidence regarding the quality differences between the models of both cell types is provided by
\autoref{fig_html_err}. It shows how the HTML error rate per tag follows the trend of the validation loss
and highlights how small differences has a large effect on the HTML quality. 
The high spread of the 6-layer LSTM HTML error rate reflect the large standard deviation observed during
training.

\subsubsection{Test cases with 128 HTML-tags}
\begin{figure*}[th!]
	\centering
	\subfloat[128 HTML-tags per case\label{fig_abs_128}]{\includegraphics[width=.5\textwidth]{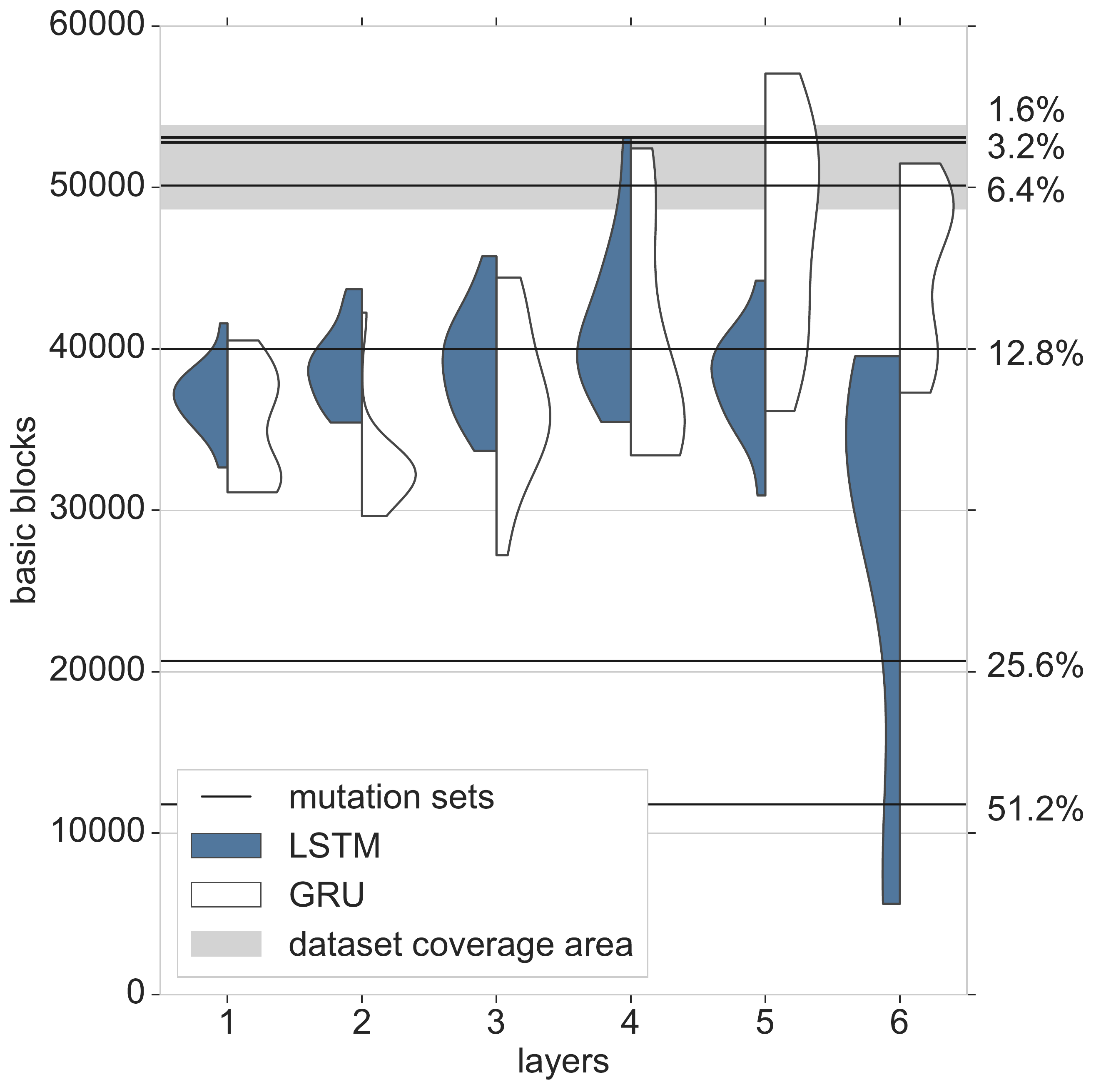}}
	\subfloat[256 HTML-tags per case\label{fig_abs_256}]{\includegraphics[width=.5\textwidth]{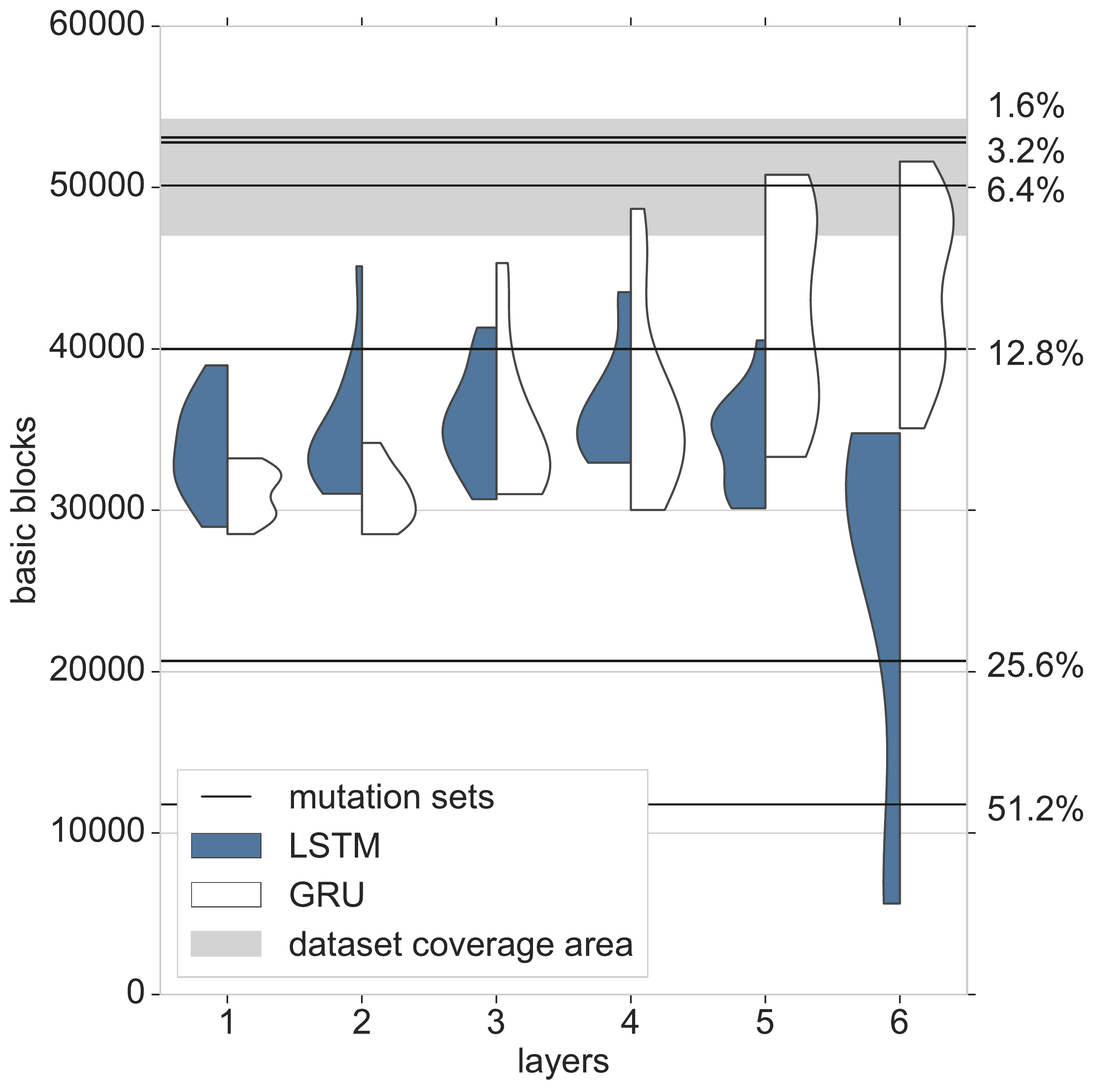}}
	\caption{Total number of uniquely discovered basic blocks on a per model basis. The dataset coverage area and the different mutation sets are included as baselines with the mutation probability indicated on the right vertical axis.}
	\label{fig_abs}
\end{figure*}

$ $\\
In terms of code coverage performance the overall trend also follows the validation loss and standard deviation,
where a smaller validation loss and standard deviation indicates a better performance. Figure \autoref{fig_abs_128} shows
the total discovered basic blocks of both cell types per layer. It highlights that both types of 4-layer models and 
the GRU 5 and 6-layer models are able to discover basic blocks in the range of the datasets or even outperform it.

In addition, Figure \autoref{fig_diff_128} shows the difference in number of basic blocks to the best performing dataset.
It shows that all models were able to discover basic blocks not triggered by the dataset, with the 5-layer GRU models
performing best on average. In comparison with the different mutation sets the maximum overlap reaches $90\%$ 
with a mutation chance of $1.6\%$, which is not surprising because the same mutation set has an overlap of
$87.6\%$ with the best performing dataset, as also shown in \autoref{fig_bb_sim}. The best performing 5-layer 
GRU models have an overlap of $78\%$ with the union of different mutation chances, highlighting the models ability 
to discover basic blocks, which can not be triggered by the naive mutation approach. The overall best performing
models are also those with the largest overlap with the dataset. 

\subsubsection{Test cases with 256 HTML-tags}
\begin{figure*}[ht!]
	\centering
	\subfloat[128 HTML-tags per case\label{fig_diff_128}]{\includegraphics[width=.5\textwidth]{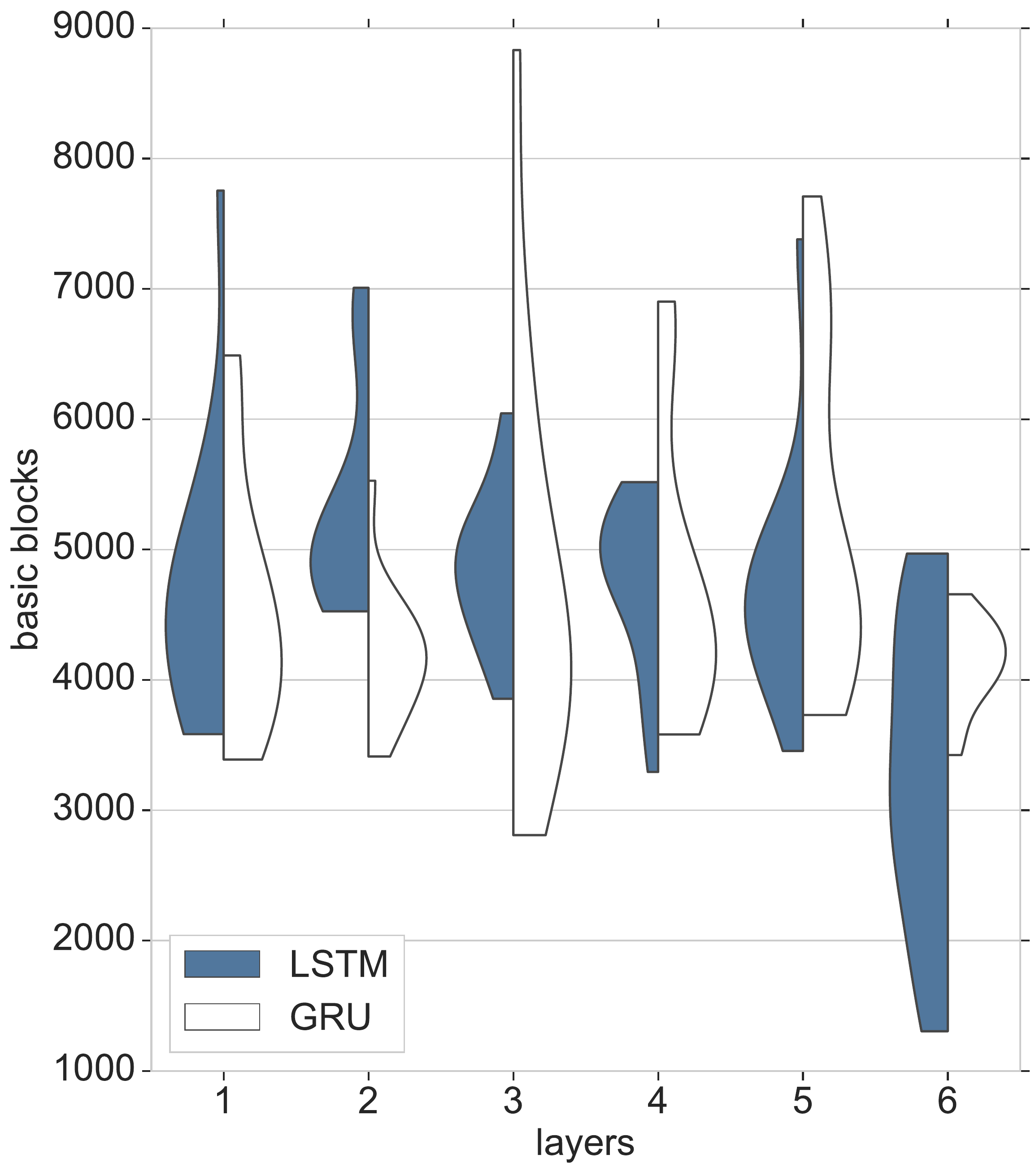}}
	\subfloat[256 HTML-tags per case\label{fig_diff_256}]{\includegraphics[width=.5\textwidth]{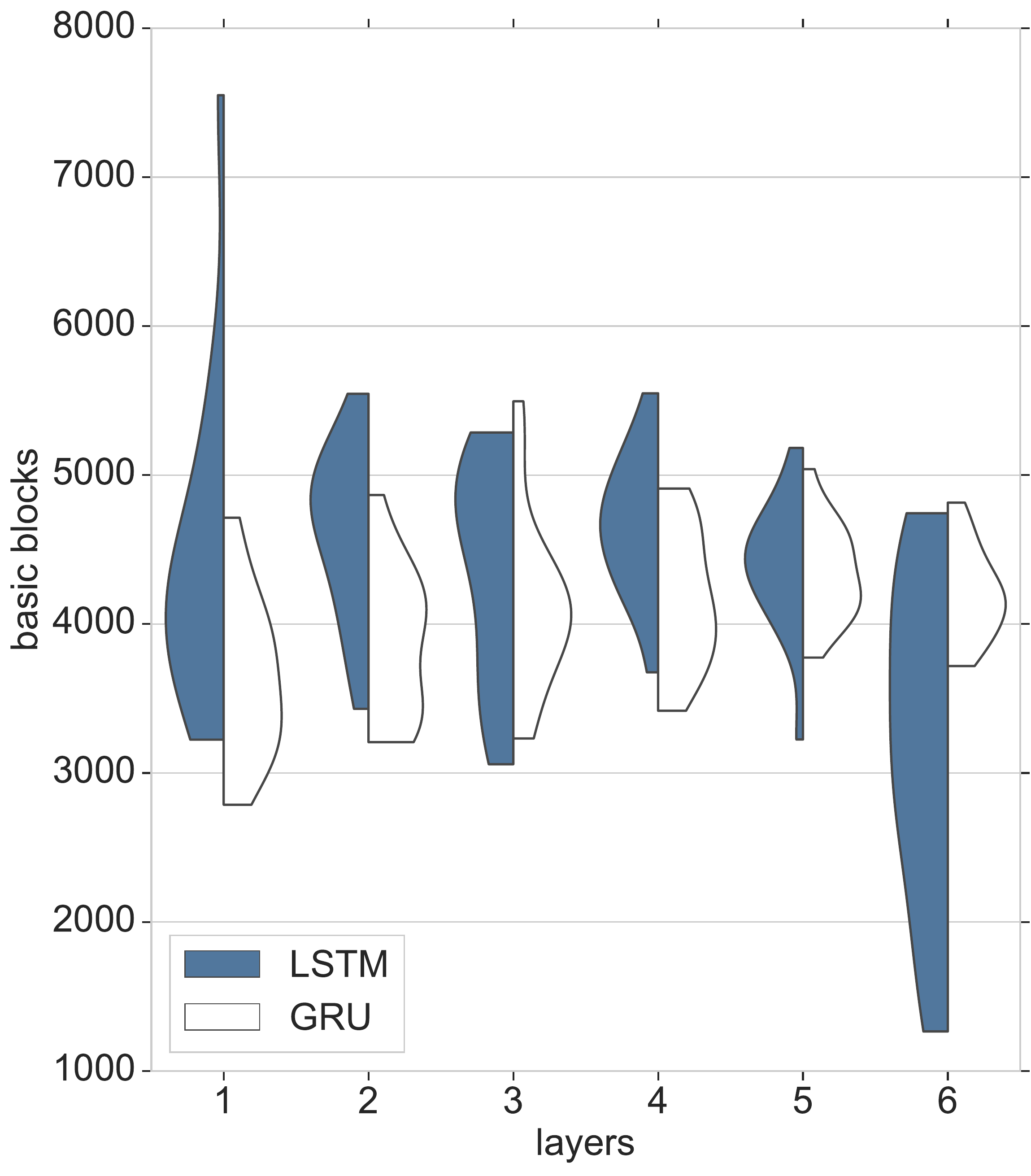}}
	\caption{Number of uniquely discovered basic blocks that were not triggered by the best performing dataset.}
	\label{fig_diff}
\end{figure*}

$ $\\
The code coverage results for the test cases with 256 HTML-tags each showed a similar development, but a slightly 
lower overall performance, as shown in Figure \autoref{fig_abs_256} and Figure \autoref{fig_diff_256}. 
The lower overall performance was expected, because both runs basically use the same HTML-tags and only the number 
of inserted HTML-tags is different. 

In terms of absolute basic blocks the 4-layer model was the best LSTM based model, however in this setting it did 
not reach the dataset coverage area. However, the 4-, 5- and 6-layer GRU based models were able to reach the dataset 
coverage area with the 6-layer model having the highest number of uniquely triggered basic blocks. 

Considering the overlap with the mutation test cases the overall result is the same as in the $128$ HTML-tags
case. The best performing four layer models have an average overlap with the mutation sets of 
$74.6\%$. This shows that the 256 HTML-tags cases were also able 
to trigger new code paths in the web rendering engine.

\section{Discussion}
The results demonstrate that is is indeed possible to successfully train models and generate test HTML cases 
using the RNN based model. However, it is crucial to monitor this process to get robust results, e.g., 
the 6-layer LSTM model was not trainable in a reliable way. This may very well have been due to a lack of 
training data, or the high amount of parameters involved in the optimisation. 

Once the models have been trained the results indicates that the average validation loss can be used
as good initial selection criteria for choosing a good model for generation of test cases despite the implicit
coupling with the code coverage metric. This is particularly interesting, since there is no code coverage data
available during the model selection phase and covering as many code paths as possible during fuzz testing is
important to discover software bugs. The results also have shown that the HTML error rate can be used to 
determine a good generative model and therefore augment the selection process. This is especially helpful,
since the average validation loss and standard deviation alone might indicate a low difference between two
models, see for example the Listings \ref{res:lst:one} and \ref{res:lst:three}. The highest average validation
loss difference between those models is $\leq 0.02$, but the difference in the HTML error rate is $0.3$. This
means that the worst performing 1-layer LSTM model has twice as many error per tag than the best performing
3-layer LSTM model.

Overall the best performing models generated more valid HTML-tags than the other models, which leads to the use
of existing HTML-tags. Those generated and generally valid HTML-tags are not always closed with right corresponding
HTML-tag. This results in the best performing models building nested valid HTML-tags by accident, because 
those models use a valid opening HTML-tag, but do not generate the corresponding closing HTML-tag. However, 
this might still be generated at a later stage in the file.
The assumed rendering behaviour and the creation of nested HTML-tags trigger code paths that have not been 
triggered by the baseline set, 
since in the baseline set every opened tag is closed with the corresponding closing tag in each line.

\begin{figure*}[t!]
	\centering
	\subfloat[LSTM\label{fig_sim_lstm}]{\includegraphics[width=.5\textwidth]{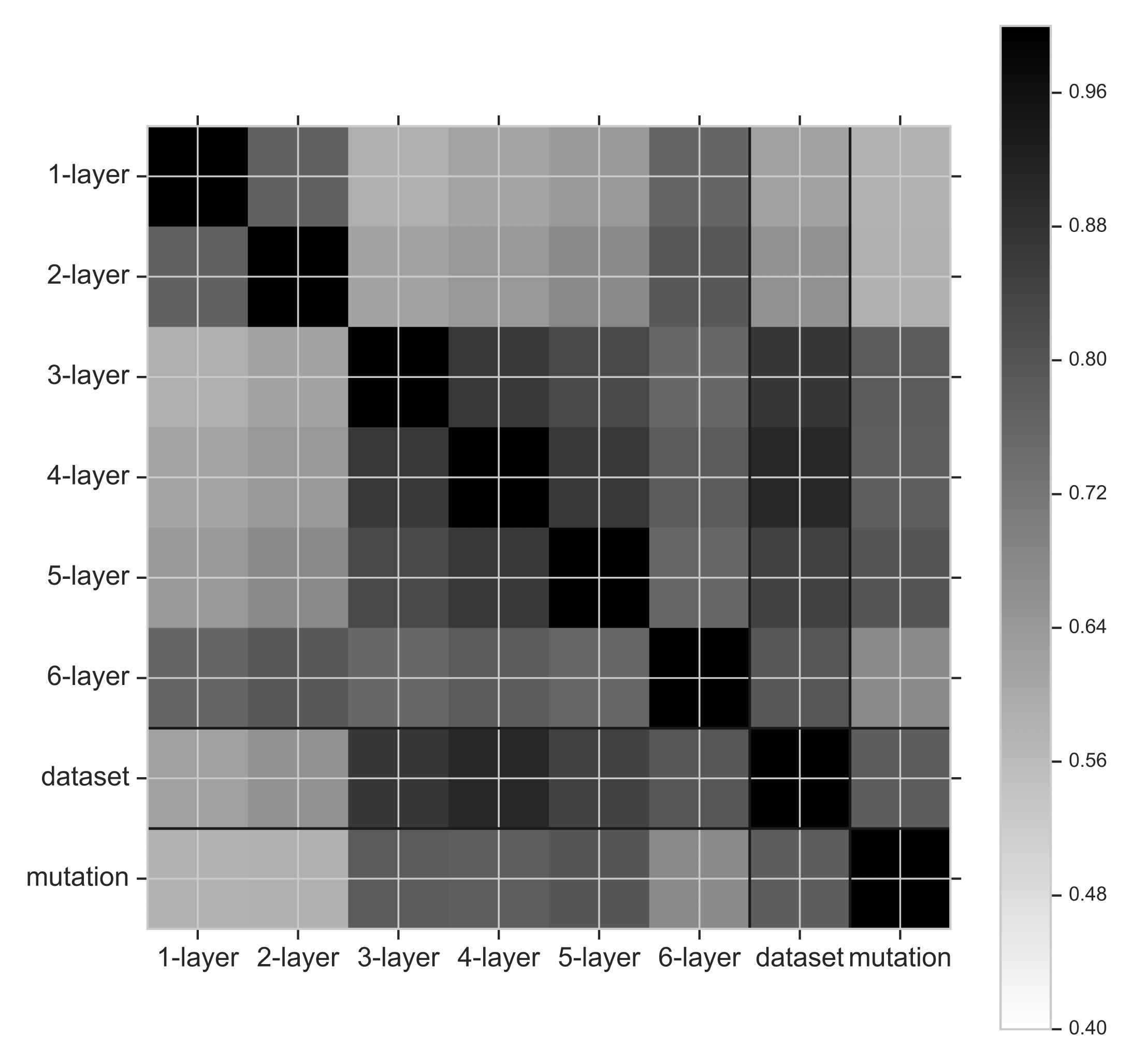}}
	\subfloat[GRU\label{fig_sim_gru}]{\includegraphics[width=.5\textwidth]{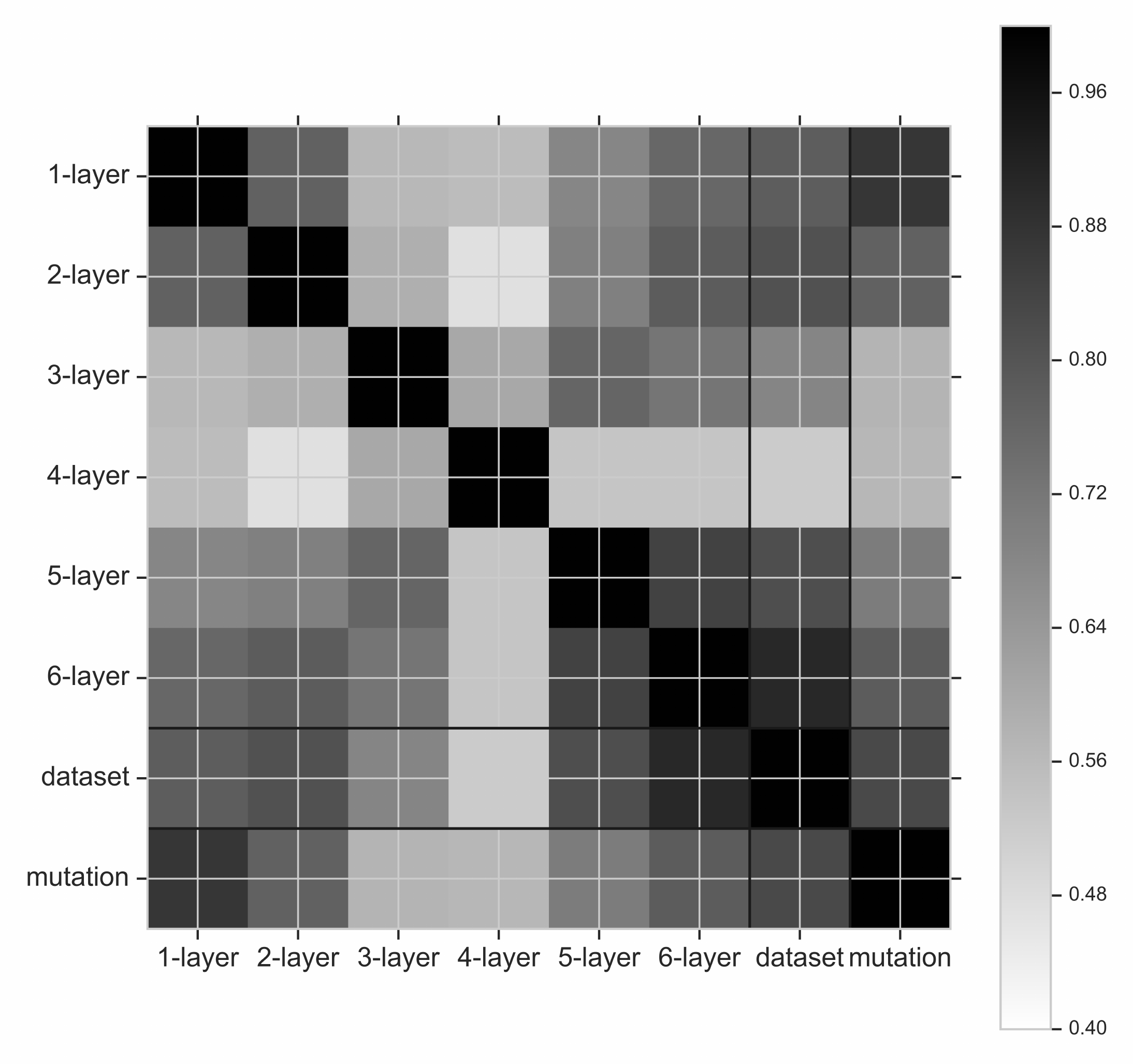}}
	\caption{The similarity between all the models, the dataset and mutation-based fuzzer in terms of their overlapping basic blocks for test cases with 128 HTML-tags.}
	\label{fig_bb_sim}
\end{figure*}

The similarity in terms of overlapping basic blocks (see Figure \autoref{fig_sim_lstm}) between the LSTM models and 
the baseline set is lower than the overlap with the mutation sets and the models between each other in
the 128 HTML-tag case. This might indicate that the models are not able to fully replicate the given input
structure and therefore another model choice would be better suited to learn this structure or the provided 
training set was too small to capture the input structure with the chosen model architecture.
For the GRU models the best performing models also show that the overlap with dataset is higher than the one
with the mutation sets (see Figure \autoref{fig_sim_gru}). This further strengthens the assumption that 
a certain quality has to be reached by the models in order perform well.

Overall, we were able to demonstrate that especially GRU-based RNNs are capable of creating HTML-tags, which then can 
be used during fuzz testing a browser. Critically, the generated HTML test cases are also able 
to trigger a significant number of unique basic blocks, which were not reached by the dataset's baseline and 
the naive mutation approach.

\section{Related Work}
The closest related work was done by Godefroid et al. \cite{godefroid2017learn}. They studied the 
achievable code coverage using a two layer stacked RNN to sample PDF-objects and focused on the effects
the training duration has on this. The code coverage results they achieved were compared against a
baseline, which was randomly selected from the training set. In contrast to that we used data not seen
by the models during the training phase to establish our baseline for comparison.
In addition, they analysed different approaches of creating test cases
and compared those. They also highlighted an observed tension between learning and fuzzing and proposed
an algorithm called SampleFuzz. This algorithm uses the lowest predicted probability, if the model's 
highest predicted probability is above a certain threshold value and a random coin toss is successful.
Whereas our work studied a different input format, namely HTML, which is a more structure-reliant input
format compared to PDF-objects. We also researched the effects of the model depth on the
resulting code coverage. We were not able to observe the former described tension between learning
and fuzzing. This might be connected to the relative large size of our training set or indicate that 
their models started to overfit to the training examples, thus requiring additional stochasticity to 
produce novel test cases. Regardless, we did not identify the need to introduce additional random values 
(e.g. through the use of SampleFuzz).

Other related works make use of the control and data flow during the execution in order to generate new
test cases. Rawat et al.\cite{rawat2017vuzzer} utilise so-called evolutionary algorithms to derive new 
test cases. Whereas H\"oschele et al.\cite{hoschele2016mining} derives an input grammar from the 
collected execution information. Both approaches need direct access to the program under test to 
instrument it and  to collect the necessary data. In contrary, our approach is able to learn the input 
structure directly from input examples, which shortens the design and learning process.

A different approach utilising code coverage and mutation-based fuzzing was presented by B\"ohme et al.
\cite{bohme2017coverage}. They augmented AFL with Markov Chains in the mutation process. Their
AFLFast called approach uses Markov Chains to determine the state transitions into new test inputs. They
have shown that they shorten the time necessary for finding bugs in an ensemble of tested software. However,
they have not provided any information on highly structure dependent input formats like HTML, which is
described as a shortfall in the general AFL approach.

Another way of combining deep learning in order to find bugs in software was evaluated by Pradel et al. 
\cite{pradel2017deep}. They used trained models in order classify potential buggy source code. Hereby
they trained their models as individual classifiers for a certain bug category. In contrast to them 
we trained our models to generate inputs, which then can be used to trigger and observe bugs in software.
Furthermore, their approach needs direct access to the source code, whereas we need access to enough 
input examples to train a RNN model.

\section{Conclusion and Future Work}
Our work provides evidence that it is possible to use a stacked RNN to generate HTML-tags in order generate novel test cases for fuzz testing a
browser's rendering engine. The results also clearly show that the GRU based models are able to outperform
LSTM ones even with less trainable parameters.  Furthermore, the proposed evaluation procedure and similarity-based analysis 
demonstrates that the overlap in basic blocks between the dataset and the model generated test cases are
very low on average. In addition, the overlap with the naively mutated sets is approximately $70\%$ on average, 
which indicates that the trained networks are able to discover new code paths formerly not discovered 
by the naive mutation approach with different mutation chances. This provides amble evidence that RNNs 
can be trained and used as an effective HTML-fuzzer provided that a suitable model-selection 
and analysis procedure is applied.

We are currently looking to extent the present work in least three ways:  
Firstly, investigating more complex/suitable neural network models is necessary to improve the overall quality of the generated HTML as other prevalent web technologies, like JavaScript, cannot be used on broken HTML-tags. Secondly, it is important to validate the generalisation of the current work on real-world HTML-examples
in contrast to the fuzzer generated training data considered here. Lastly, we are exploring ways to utilise the gathered code coverage data during the training process 
and rewarding the learning algorithm when discovering 
unintended behaviour or new code paths. We speculate that this can be achieved with the help of reinforcement learning to systematically 
trade-off the model fit vs exploration.

\paragraph{\textbf{Acknowledgements}}$ $\\
We gratefully acknowledge the support of NVIDIA Corporation with the provision of the GeForce 1080 Ti and
the GeForce TITAN Xp used for this research. We also like to thank Chris Schneider from NVIDIA for his 
ongoing interest in our research and his support.

\balance
\bibliographystyle{splncs04}
\bibliography{dissertation}

\begin{thebibliography}{10}
\providecommand{\url}[1]{\texttt{#1}}
\providecommand{\urlprefix}{URL }
\providecommand{\doi}[1]{https://doi.org/#1}

\bibitem{tensorflow2015-whitepaper}
Abadi, M., Agarwal, A., Barham, P., Brevdo, E., Chen, Z., Citro, C., Corrado,
  G.S., Davis, A., Dean, J., Devin, M., Ghemawat, S., Goodfellow, I., Harp, A.,
  Irving, G., Isard, M., Jia, Y., Jozefowicz, R., Kaiser, L., Kudlur, M.,
  Levenberg, J., Man\'{e}, D., Monga, R., Moore, S., Murray, D., Olah, C.,
  Schuster, M., Shlens, J., Steiner, B., Sutskever, I., Talwar, K., Tucker, P.,
  Vanhoucke, V., Vasudevan, V., Vi\'{e}gas, F., Vinyals, O., Warden, P.,
  Wattenberg, M., Wicke, M., Yu, Y., Zheng, X.: {TensorFlow}: Large-scale
  machine learning on heterogeneous systems (2015),
  \url{http://tensorflow.org/}, software available from tensorflow.org

\bibitem{bahdanau2014neural}
Bahdanau, D., Cho, K., Bengio, Y.: Neural machine translation by jointly
  learning to align and translate. arXiv preprint arXiv:1409.0473  (2014)

\bibitem{balog2016deepcoder}
Balog, M., Gaunt, A.L., Brockschmidt, M., Nowozin, S., Tarlow, D.: Deepcoder:
  Learning to write programs. arXiv preprint arXiv:1611.01989  (2016)

\bibitem{bengio1994learning}
Bengio, Y., Simard, P., Frasconi, P.: Learning long-term dependencies with
  gradient descent is difficult. IEEE Transactions On Neural Networks
  \textbf{5}(2),  157--166 (1994)

\bibitem{bohme2017coverage}
B{\"o}hme, M., Pham, V.T., Roychoudhury, A.: Coverage-based greybox fuzzing as
  markov chain. IEEE Transactions on Software Engineering  (2017)

\bibitem{cho2014learning}
Cho, K., Van~Merri{\"e}nboer, B., Gulcehre, C., Bahdanau, D., Bougares, F.,
  Schwenk, H., Bengio, Y.: Learning phrase representations using rnn
  encoder-decoder for statistical machine translation. arXiv preprint
  arXiv:1406.1078  (2014)

\bibitem{chung2014empirical}
Chung, J., Gulcehre, C., Cho, K., Bengio, Y.: Empirical evaluation of gated
  recurrent neural networks on sequence modeling. arXiv preprint
  arXiv:1412.3555  (2014)

\bibitem{demott2006evolving}
DeMott, J.: The evolving art of fuzzing. DEF CON  \textbf{14} (2006)

\bibitem{dynamoRIO}
DynamoRIO: Dynamorio. \url{http://dynamorio.org/} (June 2017)

\bibitem{glorot2010understanding}
Glorot, X., Bengio, Y.: Understanding the difficulty of training deep
  feedforward neural networks. In: Proceedings of the thirteenth international
  conference on artificial intelligence and statistics. pp. 249--256 (2010)

\bibitem{godefroid2017learn}
Godefroid, P., Peleg, H., Singh, R.: Learn\&fuzz: Machine learning for input
  fuzzing. Automated Software Engineering (ASE 2017)  (2017)

\bibitem{google_clusterfuzz}
Google: Using clusterfuzz.
  \url{http://dev.chromium.org/Home/chromium-security/bugs/using-clusterfuzz}

\bibitem{hochreiter1991untersuchungen}
Hochreiter, S.: Untersuchungen zu dynamischen neuronalen netzen. Diploma,
  Technische Universit{\"a}t M{\"u}nchen  \textbf{91} (1991)

\bibitem{hochreiter1997long}
Hochreiter, S., Schmidhuber, J.: Long short-term memory. Neural computation
  \textbf{9}(8),  1735--1780 (1997)

\bibitem{hoschele2016mining}
H{\"o}schele, M., Zeller, A.: Mining input grammars from dynamic taints. In:
  Proceedings of the 31st IEEE/ACM International Conference on Automated
  Software Engineering. pp. 720--725. ACM (2016)

\bibitem{ftp_rfc}
J.~Postel, J.R.: {File Transfer Protocol}. Tech. rep. (October 1985),
  \url{https://tools.ietf.org/html/rfc959}

\bibitem{jozefowicz2015empirical}
Jozefowicz, R., Zaremba, W., Sutskever, I.: An empirical exploration of
  recurrent network architectures. In: International Conference on Machine
  Learning. pp. 2342--2350 (2015)

\bibitem{kingma2014adam}
Kingma, D., Ba, J.: Adam: A method for stochastic optimization. arXiv preprint
  arXiv:1412.6980  (2014)

\bibitem{moz_ff}
{Mozilla Corporation}: Firefox. \url{https://www.mozilla.org/en-US/firefox/}
  (August 2018)

\bibitem{oehlert2005violating}
Oehlert, P.: Violating assumptions with fuzzing. IEEE Security \& Privacy
  \textbf{3}(2),  58--62 (2005)

\bibitem{pascanu2013construct}
Pascanu, R., Gulcehre, C., Cho, K., Bengio, Y.: How to construct deep recurrent
  neural networks. arXiv preprint arXiv:1312.6026  (2013)

\bibitem{pradel2017deep}
Pradel, M., Sen, K.: Deep learning to find bugs  (2017)

\bibitem{rawat2017vuzzer}
Rawat, S., Jain, V., Kumar, A., Cojocar, L., Giuffrida, C., Bos, H.: Vuzzer:
  Application-aware evolutionary fuzzing. In: Proceedings of the Network and
  Distributed System Security Symposium (NDSS) (2017)

\bibitem{sablotny_pyfuzz2}
Sablotny, M.: Pyfuzz2 - fuzzing framework.
  \url{https://github.com/susperius/PyFuzz2} (2017)

\bibitem{srivastava2014dropout}
Srivastava, N., Hinton, G., Krizhevsky, A., Sutskever, I., Salakhutdinov, R.:
  Dropout: a simple way to prevent neural networks from overfitting. The
  Journal of Machine Learning Research  \textbf{15}(1),  1929--1958 (2014)

\bibitem{sutskever2011generating}
Sutskever, I., Martens, J., Hinton, G.E.: Generating text with recurrent neural
  networks. In: Proceedings of the 28th International Conference on Machine
  Learning (ICML-11). pp. 1017--1024 (2011)

\bibitem{sutton2007fuzzing}
Sutton, M., Greene, A., Amini, P.: Fuzzing: brute force vulnerability
  discovery. Pearson Education (2007)

\bibitem{afl}
Zalewski, M.: American fuzzy lop. \url{http://lcamtuf.coredump.cx/afl/} (2017)

\end{thebibliography}

\end{document}